\documentclass[twocolumn,showpacs,showkeys,preprintnumbers,amsmath,aps,amssymb,epsfig]{revtex4-1}
\usepackage[english]{babel}
\usepackage[utf8]{inputenc}
\usepackage[T1]{fontenc}
\usepackage{textcomp}
\usepackage{graphicx}
\usepackage{bm}
\usepackage{verbatim}
\usepackage{hyperref}
\usepackage{xcolor}
\usepackage{soul}
\usepackage{cleveref}
\usepackage{ragged2e}

\begin{document}

\title{Quantum Fluids in Thermodynamic Geometry}

\author{L. F. Escamilla-Herrera} \email{lescamilla@fisica.ugto.mx}
\affiliation{Divisi\'on de Ciencias e Ingenier\'ias Campus Le\'on,
    Universidad de Guanajuato, A.P. E-143, C.P. 37150, Le\'on, Guanajuato, M\'exico.}
    \author{J. L. L\'opez-Pic\'on}
\email{jl\_lopez@fisica.ugto.mx}
\affiliation{Divisi\'on de Ciencias e Ingenier\'ias Campus Le\'on,
    Universidad de Guanajuato, A.P. E-143, C.P. 37150, Le\'on, Guanajuato, M\'exico.}
\author{Jos\'e Torres-Arenas} \email{jtorres@fisica.ugto.mx}
\affiliation{Divisi\'on de Ciencias e Ingenier\'ias Campus Le\'on,
    Universidad de Guanajuato, A.P. E-143, C.P. 37150, Le\'on, Guanajuato, M\'exico.}
    \author{Alejandro Gil-Villegas} \email{gil@fisica.ugto.mx}
\affiliation{Divisi\'on de Ciencias e Ingenier\'ias Campus Le\'on,
    Universidad de Guanajuato, A.P. E-143, C.P. 37150, Le\'on, Guanajuato, M\'exico.}

\date{\today}

\begin{abstract}

In this work,  the Thermodynamic Geometry (TG) of quantum fluids (QF) is analyzed. We present results for two models. The  first one is a quantum hard-sphere fluid (QHS) whose Helmholtz free energy is obtained from  Path Integrals Monte Carlo simulations (PIMC).  It is found that due to quantum contributions in the thermodynamic potential, the anomaly  found in TG for the classical hard-sphere fluid related to the sign of the scalar curvature, is now avoided in a considerable region of the thermodynamic space. The second model is a semi-classical square-well fluid (QSW), described by a quantum hard-sphere repulsive interaction coupled with  a classical attractive square-well contribution. Behavior of the semi-classical curvature scalar as a function of the thermal de Broglie wavelength $\lambda_B$ is analyzed for several attractive-potential ranges, and description of the semi-classical R-Widom lines defined by the maxima of the curvature scalar, are also obtained and compared with classical results for different square-well ranges.

 \end{abstract}
 
\keywords{Statistical Mechanics, Quantum Hard-Spheres, Thermodynamic Geometry, Square-well fluid}
\pacs{05.20.Jj, 95.30.Sf, 95.30.Tg}

\maketitle


\section{Introduction}


Quantum fluids are an important subject within the molecular liquids research community. Unlike classical fluids, in the quantum regime the dual wave-particle nature of the individual constituents becomes dominant and manifests in their collective properties. These systems defy our classical understanding of matter and offer exciting prospects for technological advancements, considering substances like hydrogen and helium, and their transport and storage. Additionally, the quantum nature of the hydrogen bond has been of increasing interest recently\cite{Michaelides2011} as well as its consequence in the description of the phase diagram of associating fluids \cite{Contreras2020}. On the other hand, research on QF gives insight into the fundamental principles of quantum mechanics, shedding light on the nature of matter and its interactions under extreme conditions. In this paper, we explore some aspects of the thermodynamic properties of QF using the Thermodynamic Geometry formalism, highlighting the main aspects such as the Riemann scalar curvature, and the R-Widom line.


Quantum hard spheres fluid is a natural system to study, since it is possible to compare quantum effects  with respect to the very well-known classical hard spheres system (HS). Analytical results are known for the Slater sum of QHS at low densities, obtaining  interesting insights of quantum effects in semi-classical computer simulation studies \cite{Yoon1988}; Thermodynamic, structural and dynamic properties of QHS have been studied for a wide range of densities, temperatures and thermal de Broglie wavelength values using the more robust Path-Integral Monte Carlo approach \cite{Chester1988, Sese1991, Berne2011}. Besides, in order to consider attractive interactions for a more realistic description of real fluid systems, different hard-core potential equations of state have been developed, such as for quantum square-well \cite{Singh1978,serna2016molecular} and quantum Yukawa \cite{SinghPRA1978} fluids.

In the past few decades, several approaches have been developed in order to describe thermodynamic properties of systems using differential geometry. Most of these proposals start at the well known fluctuation theory \cite{Rao1945,Mrugala1984,Mrugala1990}, assuming the existence of the notion of distance between a pair of points, each one representing a thermodynamic state, located on the space composed by thermodynamic parameters; in the context of Riemannian geometry, such a distance plays the role of the metric. In particular, Hessian metrics were postulated by Weinhold \cite{Weinhold1975} and Ruppeiner \cite{Ruppeiner1979,Ruppeiner1995}; both proposed a metric defined by the Hessian of a given thermodynamic potential in the thermodynamic equilibrium space. The Weinhold case is constructed using the Hessian of the internal energy, while Ruppeiner constructed its metric with the negative of the Hessian of the entropy; it was later proved that both metrics are related by a conformal factor \cite{Salamon1984}. The theoretical framework of such Hessian metrics is usually referred in the literature as Thermodynamic Geometry. This formalism will be considered in this work to study the physical properties of the QHS fluid.

In this work we are also interested in the study of the supercritical region of quantum systems using the tools of TG, an aspect which is of great interest for both theoretical and experimental research, and particularly the estimation of the existing boundary which separates the gas-like and liquid-like phases, which has been observed in the supercritical region \cite{Fisher1969,Widom,Bolmatov2015}. One of these proposed boundaries is the so-called Widom line \cite{Widom,Gallo2015,Jaramillo2022}, which is defined as the locus of  points that maximize the correlation length. It is known that the correlation length  diverges at the critical point, but in the vicinity above it, changes with a defined power law \cite{Fisher1969}. This behavior makes possible to characterize this curve through extreme values of the correlation length, and near to the critical point, via the extreme values of different response functions. It is well known that such curves exhibit a lineal behavior in the region close to the critical point, to later separate from each other as temperature increases. Therefore, coincidence of such lines in the $(P,T)$ plane has been used previously in the literature as a definition for a unique Widom line \cite{Widom,Brazhkin2011,Zeron2019}.

A more recent proposal to define the Widom line comes from the TG framework, that defines this curve as the locus of maxima of the isotherms of the scalar curvature obtained for the thermodynamic metric constructed using this formalism \cite{Ruppeiner2012, May2012}. In this work, this definition for the Widom line will be used to explore the supercritical behavior of the QHS fluid.

This work is organized as follows. In section II, we briefly present how  the metric and curvature are calculated in terms of the dimensionless Helmholtz free energy representation. Subsequently, equations of state related to the classical and quantum HS systems are presented together with the geometrical analysis of their curvatures, as well as a discussion about its consequences. Section III contains the same geometrical analysis for the semi-classical square-well fluid and includes for this case the Widom lines, for different values of the SW range $\lambda$. In section IV, we present the general results and discussions about the comparison with classical and quantum models. Finally, in section conclusions and future perspectives of this research are drawn and discussed.   


\section{Thermodynamic Geometry: Classical and quantum hard sphere fluid}


In this section we will consider the TG formalism \cite{Ruppeiner1979,Ruppeiner1995} for a QHS fluid. We use the Helmholtz free energy representation for which the thermodynamic metric components are obtained as second derivatives of the free energy per volume $f = A/V $ with respect to absolute temperature $T$ and  number density $\rho = N/V $. This metric is given in its matrix form by,

\begin{equation}\label{TGmetric}
[g_{ij}] = \frac{1}{kT}\ \left(
\begin{array}{cc}
-\frac{\partial^2 f}{\partial T^2}     &  0\\
0     & \frac{\partial^2 f}{\partial \rho} 
\end{array} \right)\,.
\end{equation}

In the SAFT-VR approach \cite{Gil-Villegas1997}, which will be followed in subsequent sections, the dimensionless free energy is defined by 
\begin{align}\label{Helmholtz}
\frac{A}{NkT} = a(T,\rho)\,.
\end{align}

The TG metric components of the system are determined by
\begin{align} \label{metric}
g_{_{TT}} = & \frac{6}{\pi\sigma^3} \left( \frac{2\eta}{T}\frac{\partial a}{\partial T}  - \frac{\partial^2 a}{\partial T^2} \right)\,, \\ \nonumber
g_{\eta \eta} = & \frac{\pi \sigma^3}{6} \left(  \eta \frac{\partial^2 a }{\partial \eta^2}  +2\frac{\partial a}{\partial \eta} \right)\,;
\end{align}
where $\eta$ is the packing fraction for hard-spheres particles with diameter $\sigma$, obtained from the density number as $\eta = \pi\rho\sigma^3/6$.

The corresponding reduced curvature scalar,  $R^{*} = R/\sigma^3$,  is given by
\begin{align}
R^{*} = \frac{\pi^2 \sigma^3}{36 \sqrt{g}} \frac{\partial}{\partial \eta} \left(  \frac{1}{\sqrt{g}} \frac{\partial g_{_{TT}}}{\partial \eta} \right) -
\frac{1}{\sigma^3\sqrt{g}} \frac{\partial}{\partial T} \left( \frac{1}{\sqrt{g}} \frac{\partial g_{\eta \eta}}{\partial T} \right)\,.
\end{align}

This form of $R^*$ is particularly simple because the metric is diagonal in the Helmholtz free energy representation \cite{Ruppeiner1979,Ruppeiner2012}.
According to the TG approach, negative (positive) values for $R$ indicates that attractive (repulsive) interactions are predominant on the system under study. However, it is well known that the TG description for a classical HS fluid presents a fundamental inconsistency related to the  interpretation of the sign of $R$ (see for instance Ref. \cite{Ruppeiner2012}), since the corresponding scalar curvature, calculated for different equations of state, is always negative, which contradicts the repulsive nature of the hard sphere potential.\\


In the fluid region of the phase diagram of a classical hard spheres fluid, the Helmholtz free energy can be accurately described by  the Carnahan and Starling equation\cite{Carnahan1970},  
\begin{equation}\label{CS}
\frac{A_{HS}}{NkT} =   \ln{(\rho\lambda_B^3)} -1  + \frac{4\eta - 3\eta^2}{(1-\eta)^2}\,
\end{equation}
where $\lambda_B$ is the thermal de Broglie wavelength, given by
\begin{equation}
\lambda_{B} = h/\sqrt{2\pi m kT}\,,
\end{equation}


The TG applied to the classical HS fluid was partially analyzed in Ref. \cite{Jaramillo2020}  where the curvature $R$ of the free energy \eqref{CS} was computed and compared with $R$ values determined with another HS model \cite{Ruppeiner2012}. It was observed for both models that the interaction hypothesis related to the thermodynamic curvature of the HS system seemed to be invalidated, since the HS curvature remains always negative. Given the fact that there is not an attractive contribution to the HS interaction, then the sign of its curvature contradicts the aforementioned hypothesis. This is known in the literature as the BPH anomaly \cite{BPH}. In what follows, a model of a QHS will be analyzed in order to gain insight into a possible quantum formulation of TG and with the premise that the BPH anomaly could be solved when quantum contributions are taken into account.



Quantum corrections to classical equations of state (EOS) can be introduced according to semi-classical thermodynamic perturbation methods \cite{serna2016molecular,contreras2020wertheim}. These EOS are given  as functions of the thermal de Broglie wavelength $\lambda_B$. In \cite{serna2016molecular}, the functional expression of the Carnahan-Starling excess Helmholtz free energy was used to construct a parametrization for a QHS fluid, fitting the expression in order to reproduce Path Integral Monte Carlo (PIMC) simulation values for this system.

The corresponding quantum expression for the QHS fluid can be expressed in terms of an effective temperature dependent packing fraction $\eta_e$,
\begin{equation}\label{QCSEOS}
\frac{A^{QHS}}{NkT} = \frac{4\eta_{e} - 3\eta_{e}^2}{(1-\eta_{e}^2)}\,;
\end{equation}
where $\eta_{e}$ is a function of the packing fraction and the reduced de Broglie thermal wavelength $\lambda_B^* = \lambda_{B}/\sigma$, 

\begin{equation}\label{QCS}
\eta_{e} = \left(1+d_1\lambda_{B}^{*}\right)\eta + \left(d_2 \lambda_{B}^{*} + d_3 \lambda_{B}^{*2}\right)\eta^2\,,
\end{equation}
with $d_1 = 1.6593854484$, $d_2 = -1.0927115150$, $d_3 = -1.1188233921$.

For the purpose to explore geometric properties of Eq. \eqref{QCSEOS} it is convenient to construct the TG formulation  in terms of $\lambda_{B}$ instead of  the usual variable temperature $T$.  In order to do this, it is useful to introduce the reduced temperature $T^{*} = kT/ \epsilon$, where the energy parameter $\epsilon$ is related to the potential depth of a particular attractive interaction. Although for the HS system  there is not such additional interaction, it is possible to define a suitable energy parameter, that will be denoted as $\epsilon_{0}$, given by
\begin{equation}
\epsilon_{0} = \frac{h^2}{2 \pi m \sigma^2}\,.
\label{epsilon0}
\end{equation}   
and then $T^{*}_{_{HS}} = kT / \epsilon_{0}$. 

The reduced thermal wavelength $\lambda_{B}^{*} $ can also be written in terms of the de Boer parameter $\Lambda = h/\sigma \sqrt{m \epsilon}$ in the following way,
\begin{equation}\label{deBroglie}
\lambda_{B}^{*} = \frac{\Lambda}{\sqrt{2 \pi} \sqrt{T^{*}}},  \quad T^{*} = \frac{k T}{\epsilon}\,;
\end{equation}
for the QHS system the corresponding expression is,
\begin{equation}\label{lambdaB}
\lambda_{B}^{*} = \frac{\Lambda_0}{\sqrt{2\pi}\sqrt{T^{*}_{_{HS}}}} = \frac{1}{\sqrt{T^{*}_{_{HS}}}}\,;
\end{equation}
where $\Lambda_{0} = \sqrt{2\pi}$. This particular choice for the de Boer parameter is close to the corresponding value for molecular Hydrogen. This procedure for $\Lambda$ will be also used to obtain the geometric properties of the semi-classical SW systems in the next section; since this simplification allows to write  geometric properties of the HS system in terms of the reduced thermal wavelength $\lambda_{B}^{*}$ by using the variable change $\lambda^{*}_{B} = 1/\sqrt{T^{*}_{_{{HS}}}}$. 

The geometric properties of the QHS system can be rewritten in terms of derivatives with respect to $\lambda^*_B$ using Eq.(\ref{lambdaB},

\begin{subequations}\label{metrica}
\begin{align}
g_{\lambda_B^* \lambda_B^*} &=  -\frac{3 \eta}{2\pi \sigma^3}\left( \lambda^{*2}_{B} \frac{\partial^2a}{\partial \lambda_{B}^{*2}} - \lambda_{B}^{*5} \frac{\partial a }{\partial \lambda_{B}^*} \right), \\ 
g_{\eta \eta} &=  \frac{\pi \sigma^3}{6} \left( \eta \frac{\partial a^2}{\partial \eta^2} + 2\frac{\partial a}{\partial \eta} \right)\,;
\end{align}
\end{subequations}
and hence the curvature scalar is given by :
\begin{equation} \label{escalarlambda}
R^{*} = \frac{\pi^2 \sigma^3}{36 \sqrt{g}} \frac{\partial}{\partial \eta} \left(  \frac{1}{\sqrt{g}} \frac{\partial g_{\lambda_B^* \lambda_B^*}}{\partial \eta} \right)-\frac{\lambda^{*3}_{B}}{4 \sigma^3\sqrt{g}} \frac{\partial}{\partial \lambda_{B}^*} \left( \frac{\lambda^{*3}_{B}}{\sqrt{g}} \frac{\partial g_{\eta \eta}}{\partial \lambda_{B}^*} \right)\,.
\end{equation}

As long as the free energy representation is used, the curvature $R^*$ given in \eqref{escalarlambda} remains the same, regardless  the particular model, therefore it will be used to calculate $R^*$ for the classical and semi-classical SW fluid described in the next section.

\begin{figure}[h!]
\centering
\includegraphics[width=0.375\textheight]{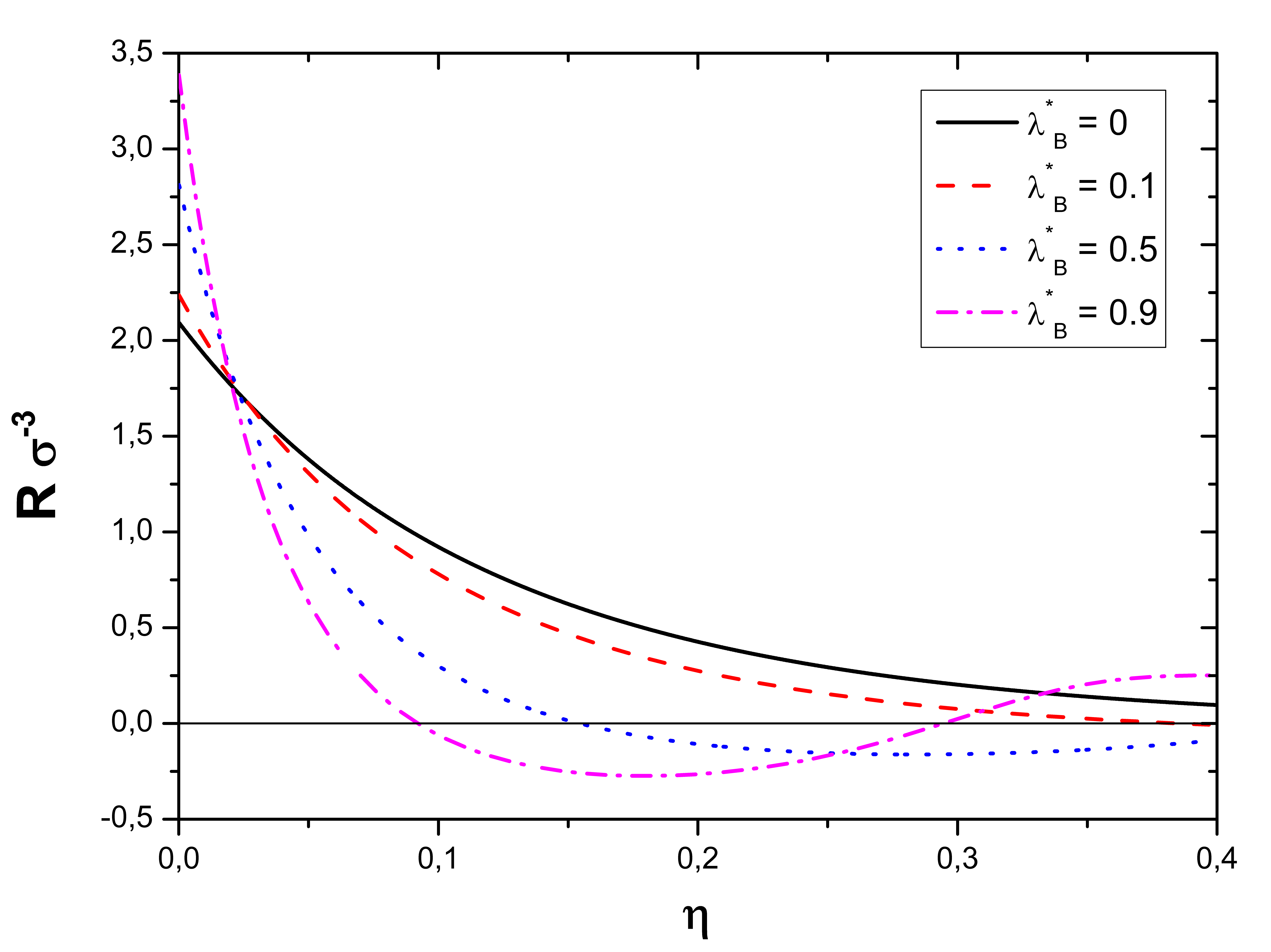}
\caption{Reduced curvature for the classical and quantum hard sphere systems for different isotherms in the ($\lambda_{B}^*$,$\eta$) representation. The solid line for $\lambda_{B}^* \to 0$ represents the behavior for a classical HS fluid; as $\lambda_{B}^*$ increases (temperature decreases), the quantum contribution becomes more important in the region where the curvature becomes negative, crossing the $R=0$ line, reversing the known anomaly for the classical HS curvature.}
\label{Rqhs2D1}
\end{figure}

Results for $R^*$ for the QHS system are presented in Fig. \ref{Rqhs2D1}.  Several isotherms in the $(\lambda_B^*,\eta)$ representation are plotted. The solid line represent the classical case (since this line correspond to the isotherm where $\lambda_B^*\to0$), which in the range of validity for the considered EOS given by \eqref{QCSEOS}, $R^*$ is always positive (i.e. negative in the $T-\eta$ thermodynamic space). As $\lambda_B^*$ increases, quantum effects become more noticeable, since a change in the sign of the curvature appears at small densities. In such region, the usual interpretation given in literature of the sign of $R^*$ is then fulfilled. However, for even greater values of $\lambda_B^*$, another change of sign appears, returning to the BPH anomaly for $R^*$ in the region of higher values of $\eta$, therefore constraining the region where quantum effects are able to revert this anomaly.
For instance, in the isotherm $\lambda_B^*=0.9$ (dash-dot-dot line) a bump underneath the line $R^*=0$ is observed, this is the region for which the usual TG interpretation of a repulsive interaction holds.

\begin{figure}[h!]
\centering
\includegraphics[width=0.375\textheight]{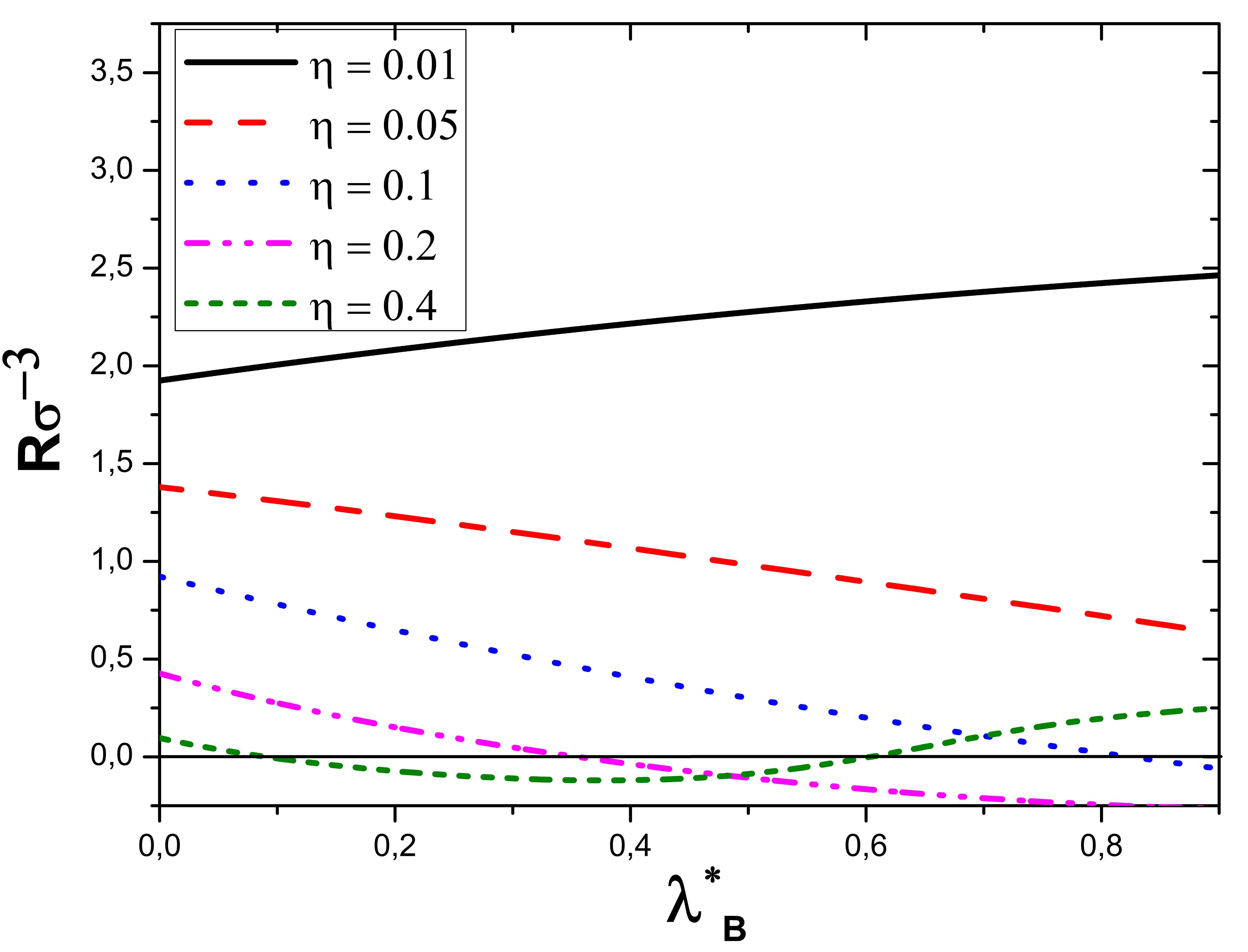}
\caption{Reduced curvature for the QHS system  in the ($\lambda_{B}^*$,$\eta$) representation for different values of $\eta$. The crossing through zero of the quantum curvature is also favored by increasing density. This behavior for both $\lambda_B^*$ and $\eta$ is better depicted in the three dimensional representation of $R^*$.}
\label{Rqhs2D2}
\end{figure}

This interpretation is reinforced in Fig. \ref{Rqhs2D2}, where $R^*$ is given in the ($\lambda_B$,$\eta$) representation for several constant values of the packing fraction. A change of sign appears in $R^*$ starting from the curve $\eta = 0.1$; the bump region where the BPH anomaly is reverted can also be noticed in this Figure, appearing in at higher densities, and it is clearly visible from $\eta=0.2$, and around the interval $0.1\le\lambda_B^*\le0.6$.

\begin{figure}[h!]
    \centering
    \includegraphics[width=0.375\textheight]{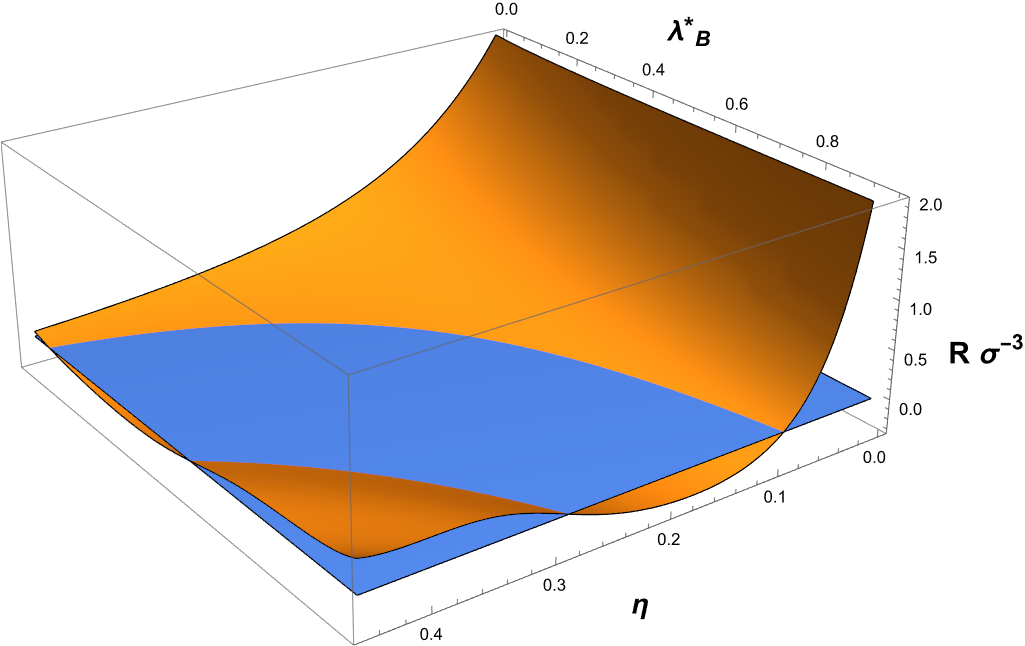}\\
        \caption{3D reduced curvature scalar for the QHS system (orange) as a function of $\lambda^*_B$ and $\eta$. The crossing with $R^* = 0$ plane (blue) is also presented. The main feature of this plot is the existence of a region of $(\lambda^*_B,\eta)$ for which the interpretation of the sign of the curvature scalar holds, located in the strip around the middle section of $\eta$.}
\label{Rqhs3D}
\end{figure}

In order to better depict the behavior of the curvature of the semi-classical QHS system, a 3D plot in the ($\lambda_{B},\eta$) representation is presented in Fig. \ref{Rqhs3D}. This plot also presents the plane of flat curvature $R^*=0$. The region below this plane with negative curvature, as observed in Fig. \ref{Rqhs2D1}, becomes a strip that cross along the middle section of the plot. This is the region where quantum effects are able to revert the BPH anomaly in the curvature of the QHS system, since in the usual $(T,\eta)$ representation, curvature becomes positive in this strip due to quantum effects, consistent with the interpretation given in the TG literature of a repulsive potential, as is the case of the HS fluid.

\begin{figure}[h!]
\centering
\includegraphics[width=0.375\textheight]{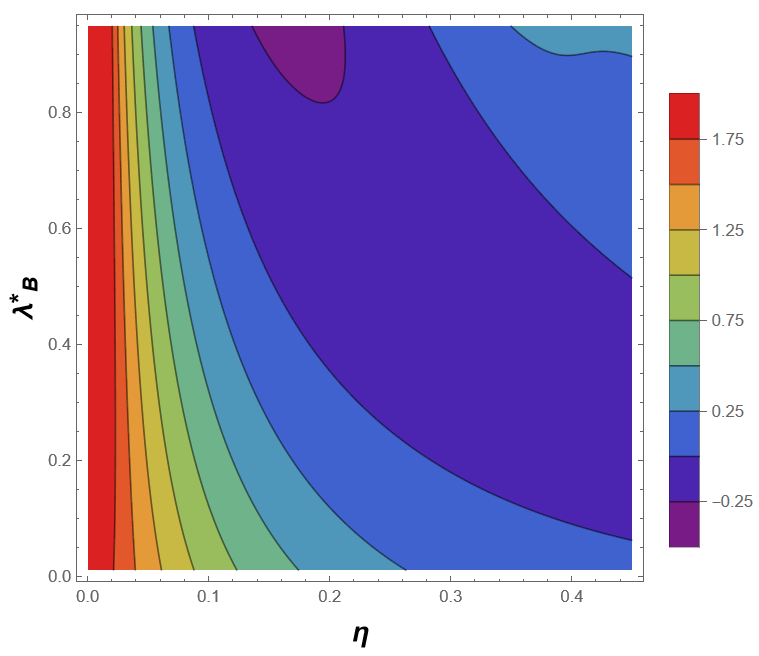}
\caption{Contour plot for the curvature scalar for the QHS system in the $(\lambda_B^*,\eta)$ space. The region where $R^*<0$ (i.e., where the BPH anomaly is reverted) is located in the strip in the middle of the plot limited by the lower and upper contour-lines (purple section), whereas for high values of $\lambda_B^*$ and $\eta$, the value of $R^*$ becomes positive once again. A region of minimum values for the curvature also appears in the upper-middle section of the plot (violet section), which can be interpreted as the region of maximal repulsion of the QHS potential.}
\label{CPRqhs}
\end{figure}

Finally, a contour plot of $\lambda_B^*$ vs $\eta$ for the QHS curvature $R^*$ is presented in Fig. \ref{CPRqhs}. This plot describes the values that curvature $R^*$ takes in the ($\lambda_B^*$, $\eta$) representation. The region where the BPH anomaly is reverted due to quantum effects is also present in the middle section, divided by the lines where $R^*=0$. This region is the same one observed in Fig. \ref{Rqhs3D}. Additionally, the existence of a small U-shaped region around $\lambda_B^* = 0.8$ and $\eta = 0.2$ can be noticed, that is the region where the lowest value (or highest value in the ($T$,$\eta$) representation) of the curvature is reached, where the QHS fluid becomes more repulsive.



\section{Thermodynamic Geometry of the semi-classical square-well fluid}

In this section,  the TG of a semi-classical SW fluid is explored. This is achieved by coupling the QHS equation of state \eqref{QCSEOS}  with a classical SW potential, a first approach to explore thermodynamic geometric properties of a quantum SW fluid. In this regard, the SW fluid is a more general and interesting system whose classical thermodynamics has been widely studied \cite{Alder1972, Henderson, Gil96, Benavides1999, Khanpour2011,Vega1992}. In the case of QSW systems, several models have been developed previously, see for instance Refs. \cite{singh1978quantum,serna2016molecular,contreras2020wertheim}. 

In this section a semi-classical analytic model is used in order to analyze its thermodynamic geometric properties via the TG framework. We consider a thermodynamic perturbation approach with  the  QHS contribution described in section II as the reference system,  given by the semi-classical Helmholtz free energy \eqref{QCSEOS}, coupled with  the classical SW contribution presented in \cite{lopez2022square} as a perturbation, 
\begin{equation}\label{QSW}
\frac{A}{Nk_{_{B}}T} = \frac{A^{\text{ideal}}}{Nk_{_{B}}T} + \frac{A^{\text{QHS}}}{Nk_{_{B}}T} + \frac{A_1}{Nk_{_{B}}T} + \frac{A_2}{Nk_{_{B}}T}\,;    
\end{equation}
where $A_{1}$ and $A_{2}$ are the first SW perturbation terms used in the SAFT-VR approach \cite{Patel2005}. The first contribution $A_{1}/NkT$ is given by
\begin{equation}
\frac{A_1}{Nk_{_{B}}T} = - 4\left(\frac{\epsilon}{kT}\right) (\lambda^{*3} -1) \eta G_{\text{HS}}(1;\eta_{\text{eff}})\,.
\end{equation}
where $G^{\text{HS}}(1;\eta_{\text{eff}})$ is the radial distribution function contact value evaluated at an effective packing fraction $\eta_{\text{eff}}$\,,
\begin{equation}
G_{\text{HS}}(1;\eta_{\text{eff}})= \frac{ 1- \eta_{\text{eff}}/2}{(1 -\eta_{\text{eff}})^3}\,.
\end{equation}

The effective packing fraction is parameterized using a Pad\'e approximation,
\begin{equation}
\eta_{\text{eff}} = \frac{c_1 \eta + c_2 \eta^2}{(1 +c_3 \eta)^3}\,,   
\end{equation}

where the indexed constants $c_i$ $(i=1,2,3)$ are given in terms of the reduced SW range, $\lambda^{*}=\lambda/\sigma$, which can be written as, 
\begin{eqnarray}
c_1&=&-\frac{3.1649}{\lambda^*}+\frac{13.3501}{\lambda^{*2}}-\frac{14.8057}{\lambda^{*3}}+\frac{5.7029}{\lambda^{*4}}\,,\\\nonumber
c_2&=&\frac{43.0042}{\lambda^*}-\frac{191.6623}{\lambda^{*2}}+\frac{273.8968}{\lambda^{*3}}-\frac{128.9334}{\lambda^{*4}}\,,\\\nonumber
c_3&=&\frac{65.0419}{\lambda^*}-\frac{266.4627}{\lambda^{*2}}+\frac{361.0431}{\lambda^{*3}}-\frac{162.6996}{\lambda^{*4}}\,.\nonumber
\end{eqnarray}

These inverse-power expansions in $\lambda^{*}$ ensure that $\eta_{\text{eff}}\to 0$ for $\lambda^* \to \infty$, from which, the desired behavior in the mean-field limit, $G_{\text{HS}} (1; \eta_{\text{eff}}) \to 1$, is recovered. 
The second-order fluctuation term $A_2/Nk_{_{B}}T$ is given by through the local-compressibility approximation \cite{Gil-Villegas1997,patel2005generalized}, 

\begin{equation}
\frac{A_2}{Nk_{_{B}}T} = \frac{1}{2} \left( \frac{\epsilon}{kT}\right) K^{\text{HS}} \eta \frac{\partial}{\partial\eta}\left(\frac{A_1}{Nk_{_{B}}T}\right).   
\end{equation}
where $K^{\text{HS}}$ is the HS isothermal compressibility \cite{lafitte2013accurate},
\begin{equation}
    K^{\text{HS}} = \frac{(1-\eta)^4}{1+4\eta + 4\eta^2 -4\eta^3+\eta^4}.
\end{equation}

Before proceeding to obtain the curvature of the free energy given in Eq. \eqref{QSW}, it is important to remark the following point, as mentioned before when the geometric properties of the QHS system were derived, it is convenient to express the free energy of the semi-classical and classical SW fluid in terms of the thermal de Broglie wavelength $\lambda^{*}_{B}$, including the explicit terms depending on temperature, according to the relation $\lambda_{B}^{*} = \Lambda/\sqrt{2\pi T^{*}}$,  with $T^{*} = kT/\epsilon$. In the case of the QHS system the energy parameter $\epsilon_0$ is used, defined in Eq. \eqref{epsilon0}. However, the SW potential already possess an energy parameter for temperature reduction the potential depth $\epsilon$.  Therefore, we have two different reduced temperatures $T^{*}_{SW}= kT/\epsilon$ and $T^{*}_{{HS}}= kT/\epsilon_0$ for the SW and HS potentials, respectively. 

From the relation between the thermal de Broglie wavelength and the de Boer parameter, presented in Eq. \eqref{deBroglie} , it follows that when this parameter is fixed to $\Lambda=\sqrt{2\pi}$, the energy parameters $\epsilon$ and $\epsilon_0$, are indeed equal, and the same is also true for the reduced temperatures. This allows us  to write the complete free energy in terms of only one reduced temperature. Besides, the relation between $\lambda^{*}_{B}$ and $T^{*}$ is simply $\lambda^{*}_{B} = 1/\sqrt{T^{*}}$ and the change of variable applied to calculate the curvature of the QHS system can also be applied without any further modifications, and the corresponding expressions for the metric components and curvature are the same than the ones given by Eq. \eqref{metrica} and Eq. \eqref{escalarlambda}, respectively.

\begin{table}[h!]
    \centering
    \begin{tabular}{| c | c | c | c | c | c | c |}
    \hline
      $\lambda^*$   & $\eta_{\text{qcrit}}$ & $T^*_{\text{qcrit}}$ &  $P^*_{\text{qcrit}}$ & $\eta_{\text{ccrit}}$  & $T^*_{\text{ccrit}}$ & $P^*_{\text{ccrit}}$ \\
    \hline
    \hline
       1.25  & 0.0314 & 0.2378 & 0.0047 & 0.2239  & 0.83627 & 0.1649 \\
       1.50  & 0.0404 & 0.4932 & 0.0126 & 0.1499  & 1.32907  & 0.1434 \\
       1.75  & 0.0475 & 0.8487 & 0.0255 & 0.1232  & 1.9073   & 0.1583  \\
       2.00  & 0.0566 & 1.3782 & 0.0493 & 0.1247  & 2.8248   & 0.2325 \\
       2.50  & 0.0783 & 3.3019 & 0.1673 & 0.1330  & 5.8015   & 0.5226 \\
       3.00  & 0.0931 & 6.7089 & 0.4201 & 0.1336  & 10.3288  & 2.3393 \\
    \hline
    \end{tabular}
    \caption{Theoretical critical values in the ($T^*,\eta$) representation of the semi-classical and classical SW fluid are presented for different potential ranges between 1.25 an 3.0, denoted by \textit{qcrit} and \textit{ccrit}, respectively.}
    \label{table1}
\end{table}

The TG metric of the semi-classical SW fluid is obtained through second-order derivatives of the Helmholtz free energy given in Eq. \eqref{QSW} and applying Eq. \eqref{TGmetric}. In order to compare how this geometric property changes by introducing the QHS potential given in Eq. \eqref{QCSEOS}, the classical SW fluid model presented in \cite{lopez2022square} is used as a reference. In Table \ref{table1} the critical values for different potential ranges for the classical and semi-classical SW fluid are presented in the $(T^*,\eta)$ representation. It can be noticed that the critical values of the semi-classical system are considerably lower when compared to the classical one. 

\begin{figure}[h!]
\centering
\includegraphics[width=0.375\textheight]{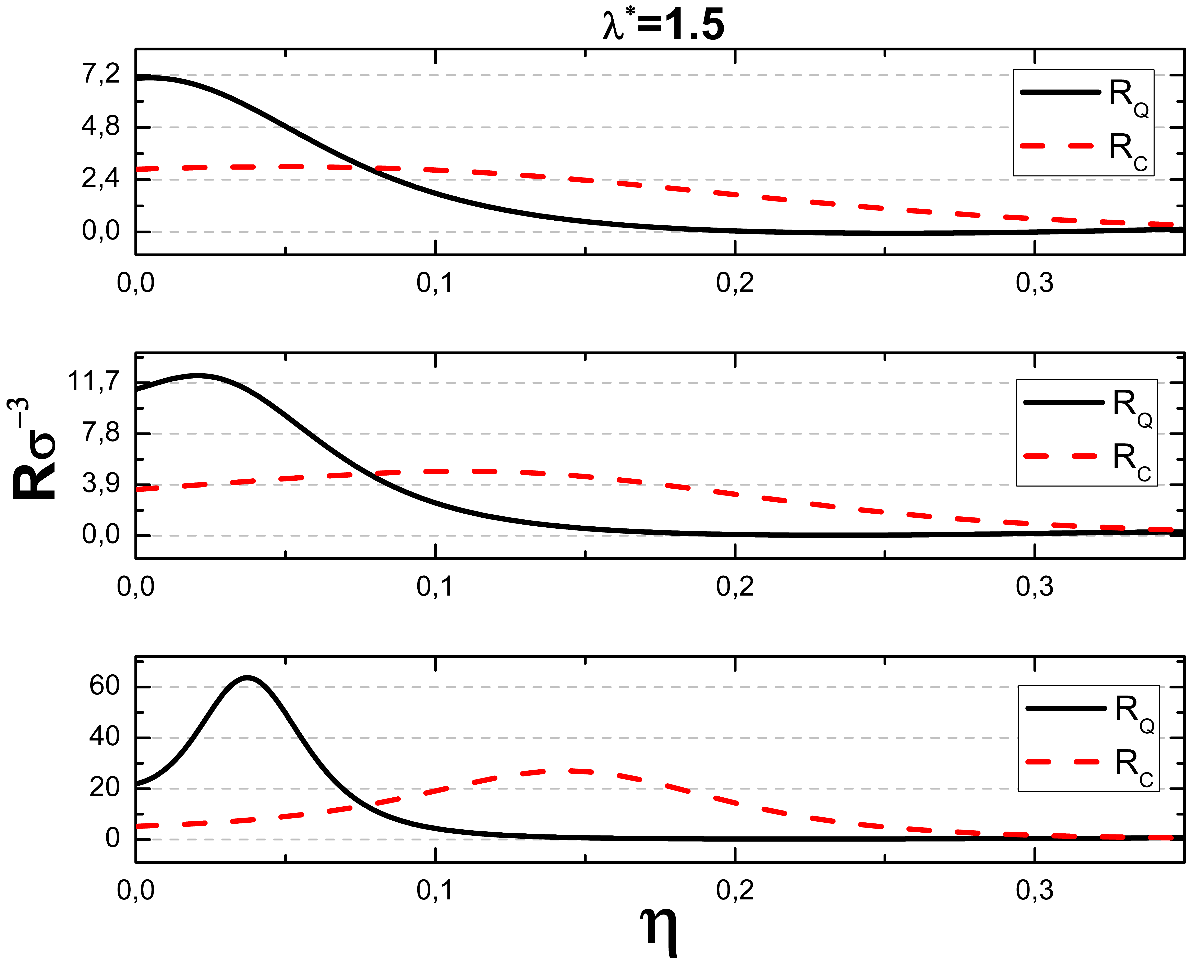}\\
\caption{Isotherms of the curvature $R$ of the semi-classical and classical SW fluid for a potential range of $\lambda^* = 1.5$ in the supercritical region are presented, given by solid and dashed lines, respectively. To have a better insight on the behavior of both systems, $T^*$ were chosen to be at $1.25 \ T_{\text{crit}}^*$ (bottom), $2 \ T_{\text{crit}}^*$ (middle) and $3 \ T_{\text{crit}}^*$ (upper), namely at the same percentage away from its corresponding critical value.}
\label{R-slides}
\end{figure}

\begin{figure}[h!]
\centering
\includegraphics[width=0.375\textheight]{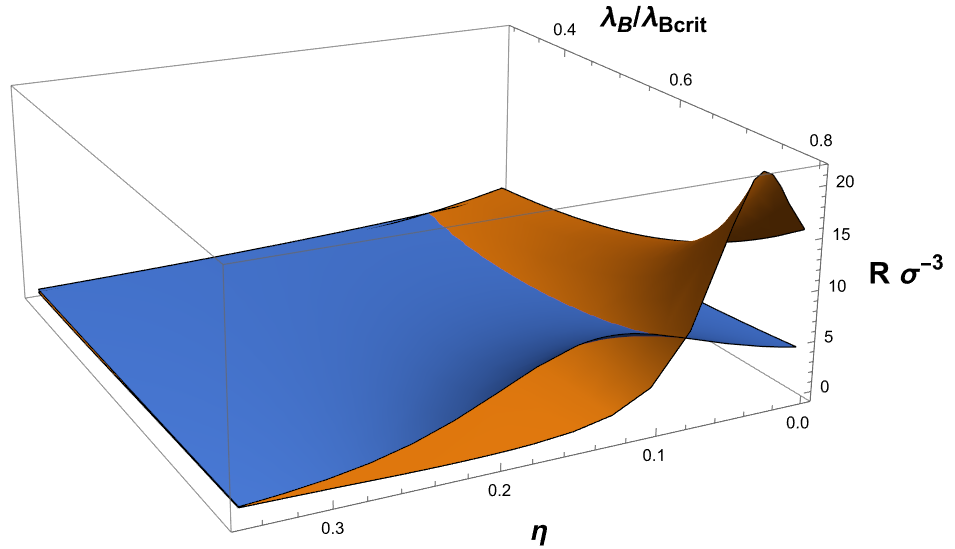}\\
\caption{3D plot of the reduced classical and semi-classical  SW curvatures, represented by blue and orange surfaces, respectively. Both plots are determined for the same SW range, $\lambda^* = 1.5$. The corresponding thermal de Broglie wavelength values at the critical temperatures are $\lambda_{B}^{SC}=1.4239$ and $\lambda_{B}^{C}= 0.8674$ for the semi-classical and classical curvatures, respectively. In order to avoid singularities due to the gas-liquid transition, the plot is constructed considering only supercritical values of $\lambda_B^*$ for each system between $0.3\le \lambda_B^*/ \lambda_{B\text{crit}}^* \le 0.8$.}
\label{3dsw}
\end{figure}

Once both metric elements and curvatures are calculated for the semi-classical and classical SW fluids using Eqs. \eqref{metric} and \eqref{escalarlambda} in the $(\lambda_B^*,\eta)$ representation, an analysis of their behavior was performed in the supercritical region of $\lambda_B^*$, in order to avoid any singularity due to the gas-liquid phase transition. First, a 2D comparison of the curvature scalar for different isotherms is presented in Fig. \ref{R-slides} for a SW range  $\lambda^* = 1.5$ for both semi-classical and classical systems,  solid and dashed lines, respectively. Three different temperatures have been considered for this comparison, namely, $T^* = 1.25 \ T_{\text{crit}}^*$, $T^* = 2.0 \ T_{\text{crit}}^*$ and $T = 3.0 \ T_{\text{crit}}^*$; these particular values were chosen to fairly compare the corresponding curvatures since the critical points for each system are not close to each other, as noticed in Table \ref{table1}. The main remark about Fig. \ref{R-slides} is that the maximum of $R^*$ reached for the semi-classical SW is higher when compared to the classical one at the same ``distance'' to their corresponding critical point; with a steeper maximum for the semi-classical system and located at smaller packing fractions compared to the classical one. This feature is more evident as reaching the critical point of each system.\\

Fig. \ref{3dsw} presents 3D plots of the semi-classical (orange surface) and classical (blue surface) SW fluid  for the curvature scalars in the thermodynamic space $(\eta,\lambda_{B}^*)$ for a SW $\lambda^* = 1.5$. As in the case of the two-dimensional plots presented in Fig. \ref{R-slides}, these surfaces were  obtained for the supercritical region to avoid any singularity in $R^*$ close to the critical point, starting from $0.3 \ \lambda_{B\text{crit}}^*$ up to $0.8 \ \lambda_{B\text{crit}}^*$, i.e. from 30 to 80 percent of the critical value in $\lambda_B^*$. From this figure it can be noticed that  the curvature of the semi-classical and classical SW fluids greatly differs in the region $\eta\to0$, i.e., for small densities, where quantum effects are more relevant. This behavior is also captured in the 2D plots.

\begin{figure}[h!]
\centering
\includegraphics[width=0.375\textheight]{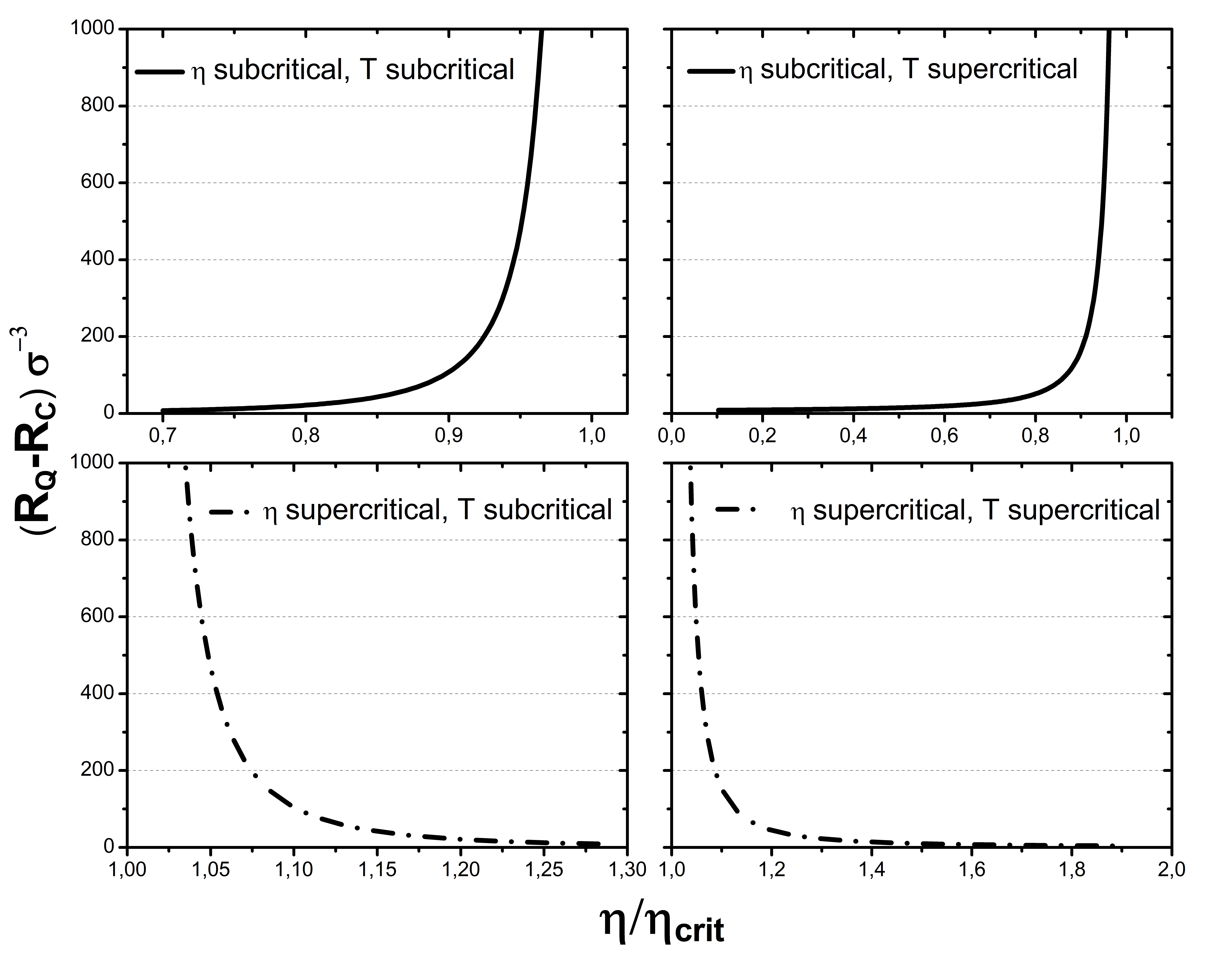}\\
\caption{Differences between classical and semi-classical SW curvatures for $\lambda^* = 1.5$. These differences were computed approaching the critical point in four different directions: (1) Both parameters in the sub-critical region (upper left); (2) approaching critical density from the sub-critical region and critical temperature from the supercritical region (upper right); (3) density in the supercritical region ant temperature approaching the critical point from the sub-critical region (bottom left);(4) both parameters in the supercritical region approaching the critical point (bottom right).}
\label{R-differences}
\end{figure}

In order to display the differences in the behavior between the semi-classical and classical curvature scalars of the SW fluid,  the difference $R^*_Q-R^*_C$ of the reduced semi-classical and classical curvatures for  $\lambda^* = 1.5$,  $R^*_Q$ and $R^*_C$, respectively, is presented in Fig. \ref{R-differences}.  As can be seen  in Fig \ref{R-differences}, each curve is always positive for the considered interval of $\eta/\eta_{\textbf{crit}}$, indicating that $R^*_Q$ is always greater than $R^*_C$.

To remark these differences, four different temperature regions are shown separately: (1) $\eta<\eta_{B\text{crit}}$ and $T^*<T_{B\text{crit}}^*$ in the upper left plot; (2) $\eta<\eta_{B\text{crit}}$ and $T^*>T_{B\text{crit}}^*$ in the upper right plot; (3) $\eta >\eta_{B\text{crit}}$ and $T^* < T_{B\text{crit}}^*$ in the bottom left; (4)  $\eta>\eta_{B\text{crit}}$, and $T^*>T_{B\text{crit}}^*$ in the bottom right side. The behavior of the differences in the supercritical region of $\eta$ is the same with $\left(R^*_Q-R^*_C\right)\to0$ as $\eta^*$ goes far away from  its critical value. This is a hint that at higher densities, geometric properties of the semi-classical system resembles those of the classical SW fluid. In the opposite way, in the subcritical and supercritical region of temperature, the difference $\left(R^*_Q-R^*_C\right)\to\infty$ as $\eta^*\to 0.9$, which represents the divergence that appears at $R$ as the critical point is reached.

\begin{figure}[h!]
\centering
\includegraphics[width=0.375\textheight]{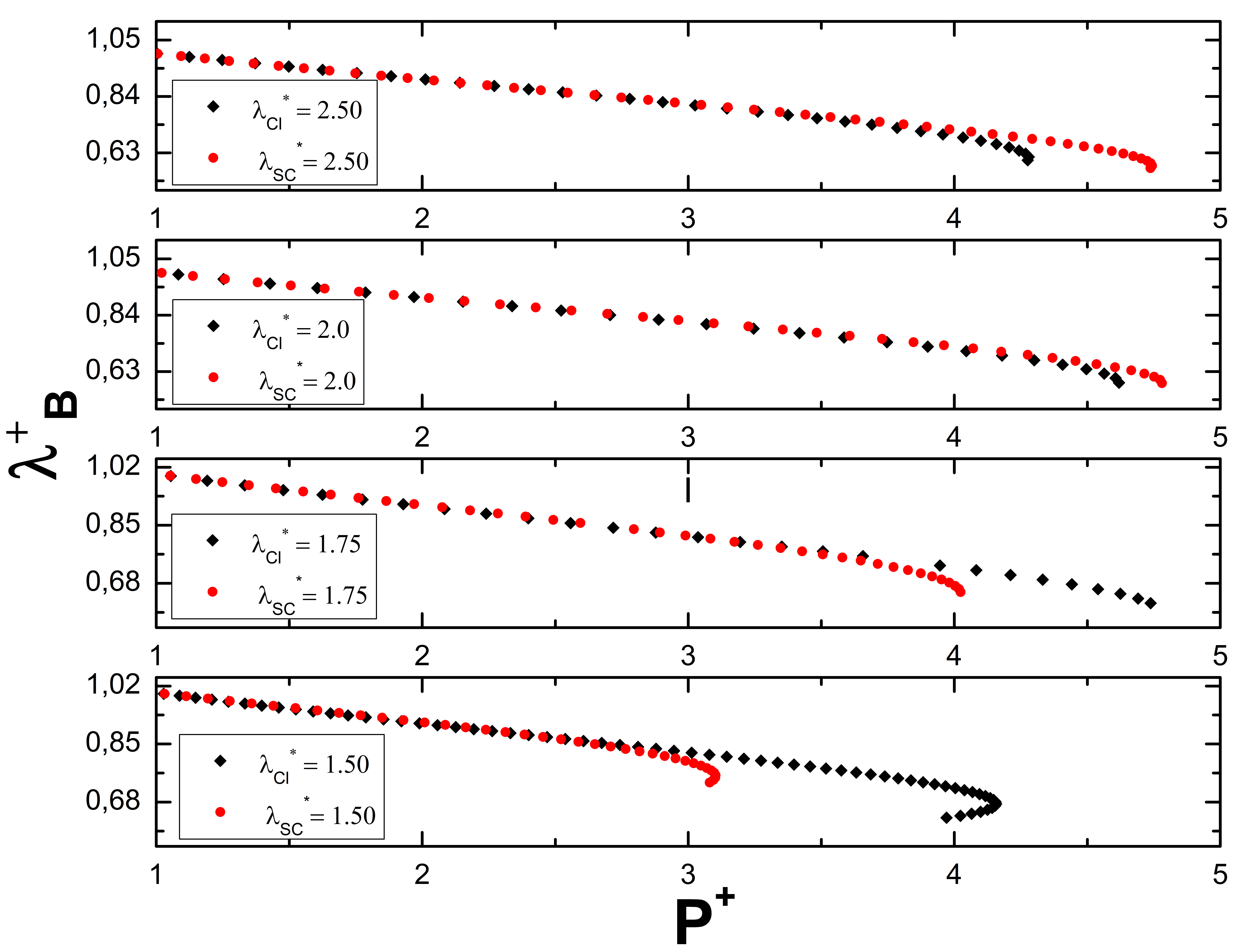}\\
\caption{Comparison of $R$-Widom lines obtained 
for the  semi-classical and classical SW fluids, given by  dots and diamonds, respectively, for the following values of the SW range  $\lambda^*$: 1.5 (bottom), 1.75 (middle bottom), 2 (middle upper), and 2.5 (upper). Results are presented as a function of the reduced de Broglie thermal wavelength $\lambda^+_B=\lambda_B/{\lambda_B}_{\text{crit}}$ and the reduced pressure $P^+=P/P_{\text{crit}}$. In this representation, higher supercritical states are reached as $\lambda_B^*\to0$. Semi-classical and classical lines exhibit the same linear behavior in the region near the critical point. The supercritical line can be extended further for the semi-classical SW fluid for smaller potential ranges. Interestingly, this behavior is reversed as $\lambda^*$ increases.}
\label{Widom}
\end{figure}

An additional comparison of the differences in the geometric properties of the semi-classical and classical SW fluid is also performed, via an analysis on the behavior of the $R$-Widom lines for both systems. As mentioned in the introduction section, these lines are defined in the supercritical region as the locus of the maxima for the isotherms of the scalar scalar curvature $R$; this line separates a gas-like from a liquid like phase in the supercritical region for the semi-classical and classical SW fluid. This comparison is presented in Fig. \ref{Widom} in the $(\lambda^+_B,P^+)$ representation, instead of the usual temperature-pressure one, therefore, separation of phases is inverted with respect to the usual case. The most relevant feature of these plots is that when semi-classical and classical Widom lines are compared, these lines are range dependent. For smaller ranges ($\lambda^* < 2.0$), a shorter Widom line is obtained for the semi-classical system. For instance, when $\lambda^* = 1.5$ the semi-classical line cuts around $P^+\approx 3$, while its classical counterpart have a linear behavior up to $P^+\approx 4$. This situation is reversed, however, as $\lambda^*$ is increased, as can be seen for $\lambda^* = 2.0$; in this case the longest Widom line is the semi-classical. Therefore, for greater values of the SW range, the line which separates the aforementioned liquid-like phase from the gas-like is more persistent for the semi-classical system.

\section{Conclusions}
In this work, a short glance of the Thermodynamic Geometry  of quantum fluids was explored. Two models different were analyzed, a quantum hard-sphere fluid whose Helmholtz free energy is obtained from Path Integrals Monte Carlo simulations and  a semi-classical square-well fluid, described by a quantum hard-sphere repulsive interaction coupled with  a classical attractive square-well contribution. 
The curvature obtained from the QHS potential presents important differences compared to its classical counterpart. 

It is found  that when quantum contributions are taken into account, the BPH anomaly is partially reverted, in a region in the middle of the plane $(\lambda_B^*,\eta)$ where the considered QHS equation of state  is valid.  For the  regions of low and high densities, $R_{\text{QHS}}$ exhibits the same anomalous behavior than its classical counterpart.  It is not surprising that at lower densities the classical anomalous behavior remains, however, it is interesting that at the regime of high densities the QHS  curvature tends to return to the anomalous classical behavior since it resembles the behavior of other quantum systems \cite{markland2011quantum}.

Regarding the geometric properties of the semi-classical SW fluid, when compared to the classical one, interesting differences were found between both systems. For instance, the critical point in temperature and density of the semi-classical system is significantly lower that the classical one, which can be noticed in the direct comparison of the supercritical curvatures depicted in Fig. \ref{3dsw}. This comparison is further explored at the difference $R^*_Q-R^*_C$, where we have found, for each of the four cases explored, that the semi-classical curvature is always greater than the classical one. Finally, the supercritical $R$-Widom lines for both systems were also determined and compared, showing that near the critical point both lines practically overlap.Interestingly, it was found that the length of these lines are also potential-range dependent, and the length behavior for ranges below $\lambda^* = 2.0$ is larger for the classical SW fluid, which is reversed for ranges above this value. Then we can distinguish the liquid-like phase from the gas-like one deeper into the supercritical region when compared to the classical one.

It is clear that this research represents only a first step in the study of quantum fluids within the framework of Thermodynamic Geometry, and  more work is needed in order to clarify the influence and consequences of quantum effects in these systems.\\

\newpage

\acknowledgments{J. Torres-Arenas  acknowledge support by University of Guanajuato through grant 174/2023 of Convocatoria Institucional de Investigaci\'on Cient\'ifica 2023. L.F. Escamilla-Herrera acknowledge support from CONAHCYT through posdoctoral grant: Estancias Posdoctorales por México para la Formación y Consolidación de las y los Investigadores por México.} 

\bibliographystyle{apsrev}
\bibliography{QuantumHS}

\end{document}